\documentclass[pop,fleqn]{w-art}
\usepackage{times}
\usepackage{w-thm}
\usepackage{amssymb,amsmath}
\usepackage[]{graphicx}
\begin{document}

\DOIsuffix{theDOIsuffix}
\Volume{51}
\Issue{1}
\Month{01}
\Year{2003}



\title[Short title]{Probe D--branes in Superconformal Field Theories}

\author{Jos\'e D. Edelstein\inst{1,2,}%
  \footnote{Corresponding author\quad E-mail:~\textsf{jedels-at-usc-dot-es}, 
            Phone: +34\,981\,563\,100 ~x.13980, 
            Fax: +34\,981\,521\,091}}
\address[\inst{1}]{Departamento de F\'\i sica de Part\'\i culas and IGFAE,
Universidade de Santiago de Compostela, E-15782 Santiago de Compostela, Spain}
\address[\inst{2}]{Centro de Estudios Cient\'\i ficos (CECS), Casilla 1469,
Valdivia, Chile}
\begin{abstract}
We overview the main configurations of D--brane probes in the $AdS_5 \times
X^5$ background of type IIB string theory ($X^5$ being a Sasaki--Einstein
manifold), and examine their most salient features from the point of view of
the dual quiver superconformal field theory.
\end{abstract}
\maketitle                   


\renewcommand{\leftmark}
{Jos\'e D. Edelstein: Probe D--branes in Superconformal Field Theories}


\section{Introduction}

The open/closed string descriptions of a system of $N_c$ parallel D3--branes
in flat space decouple in the large $N_c$ limit, this leading to the
duality between $\mathcal{N} = 4$ supersymmetric Yang--Mills theory and
type IIB string theory in $AdS_5 \times S^5$ \cite{Malda}. If, instead, the
transverse six dimensional flat space background is replaced by a Calabi--Yau
threefold, $Y^6$, the amount of preserved supercharges reduces to one quarter.
\footnote{For the sake of reducing supersymmetry (while also giving up
conformal invariance), another avenue involving higher dimensional D--branes
wrapping supersymmetric cycles of $Y^6$ has been explored \cite{MaNu} (see
\cite{EdPo} for a recent review with updated references).} Provided $Y^6$
is a cone on a Sasaki--Einstein manifold $X^5$, and the stack of
D3--branes is placed at the tip of the Calabi--Yau cone $\mathcal{C}(X^5)$,
a duality between quiver $\mathcal{N}=1$ superconformal field theories (SCFT)
and type IIB string theory in $AdS_5 \times X^5$ arises \cite{Gu}. The
case in which $Y^6$ is also toric, and $X^5$ is topologically $S^2 \times
S^3$, is by now very well understood. There are three possibilities
\footnote{It is convenient to clarify at this point that they are not
independent: $Y^{p,q}$ happens to be a subfamily of $L^{a,b,c}$
(indeed, $Y^{p,q} = L^{p-q,p+q,p}$), and --an orbifold of-- $T^{1,1}$ can be
obtained as a singular limit of $Y^{p,q}$ (meaningly, $Y^{p,0} =
T^{1,1}/\mathbb{Z}_p$).} (whose main features are displayed in Table 1):
\begin{itemize}
\item
$X^5 = T^{1,1}$: ~Its isometry is $SU(2) \times SU(2) \times U(1)$. Its metric
has been constructed long ago \cite{CadlO}. The dual gauge theory was worked
out soon after the advent of the AdS/CFT correspondence \cite{KlWi}.
\item
$X^5 = Y^{p,q}$: ~Its isometry is $SU(2) \times U(1) \times U(1)$. Its metric
has been discovered more recently \cite{metricYpq}. It depends on two positive
integers $p$ and $q$. The dual gauge theory was puzzled out in
\cite{SCFTYpq}.

\item
$X^5 = L^{a,b,c}$: ~Its isometry is $U(1) \times U(1) \times U(1)$. Its metric
was found much more recently \cite{metricLabc}. It depends on three positive
integers $a$, $b$ and $c$. The dual SCFT was unraveled in \cite{SCFTLabc}.
\end{itemize}

A crucial ingredient of the AdS/CFT duality is the state/operator
correspondence: chiral operators of the CFT are associated with supergravity
modes in the dual background.
\begin{table}[h]
\begin{center}
$$\begin{array}{|c|c|c|c|}
\hline
& & & \\[-2ex]
X^5 & \mathrm{Isometry/Global~Symmetry} & \mathrm{Bifundamental~Chiral~Fields}
& N_{gg} \\
& & & \\[-2ex]
\hline\hline
& & & \\[-1ex]
T^{1,1} & SU(2) \times \widetilde{SU}(2) \times U(1) & A^{\alpha\,[1]},
\tilde{B}^{\alpha\,[-1]} & 2 \\[1ex]
\hline
& & & \\[-1ex]
Y^{p,q} & SU(2) \times U(1) \times U(1) & U^{\alpha\,[-p]}_p,
V^{\alpha\,[q]}_q, Y^{[p-q]}_{p+q}, Z^{[p+q]}_{p-q} & 2p \\[1ex]
\hline
& & & \\[-1ex]
L^{a,b,c} & U(1) \times U(1) \times U(1) & S^{[-c]}_{a+b-c}, T^{[c-a-b]}_c,
W^{[b-c]}_{c-a}, X^{[c-a]}_{b-c}, Y^{[a]}_b, Z^{[b]}_a & a+b \\[1.5ex]
\hline
\end{array}$$
\caption{Data corresponding to the quiver $\mathcal{N}=1$ theories
under discussion. In the third column, there is an upper index $\alpha$ used
for doublet fields (with respect to the appropriate $SU(2)$ factor), there is
a subindex that indicates the degeneracy ({\it i.e.}, how many bifundamental
chiral fields have the same quantum numbers), and there is an upper label in
brackets that displays the $U(1)_B$ charge. The gauge group of each
theory is $SU(N_c) \times \dots \times SU(N_c)$, $N_{gg}$ times.}
\label{charges}
\end{center}
\end{table}
Still, there are features of the gauge theory whose description demands the
introduction of (wrapped) D--branes in the gravity side. Most notably, {\it
dibaryon operators} corresponding to each bifundamental chiral field of
these quiver gauge theories. They are given by D3--branes wrapping
supersymmetric 3-cycles in $X^5$ \cite{Wit,dibar,BeHeKl}. These are
point-like objects from the SCFT point of view. This is also the case for
the {\it baryon vertex}, corresponding to a baryon built out of external
quarks, that is represented by a D5--brane wrapping $X^5$ \cite{Wit}.  

Extended objects in the gauge theory side also correspond to wrapped
D--branes in the string theory side. String-like objects as {\it confining}
or {\it fat strings} arise from D3--branes wrapping $2$-cycles. {\it Domain
walls}, {\it fractional branes} and {\it defect CFTs} are given by D5--branes
wrapping 2-cycles in $X^5$. The introduction of {\it matter hypermultiples}
--that is, quarks in the fundamental representation--, requires spacetime
filling wrapped D7--branes \cite{KaKa}. If the number of wrapped D--branes
is much less than $N_c$, we can stick to the probe approximation. For
instance, this is the case when matter is introduced in the quenched
approximation, $N_f \ll N_c$. This is the framework considered in the
present talk.

\section{Some geometrical facts}

Let us consider a solution of IIB supergravity whose metric is of the form
$ds^2 = ds^2_{AdS_5} + ds^2_{X^5}$ (we choose, for simplicity, a unit radius
$L=1$ for both spaces). The metrics $ds^2_{X^5}$ can be locally written as
\begin{equation}
ds^2_{X^5}\,=\,ds^2_{4} + \left[ \frac{1}{3} d\psi + \sigma
\right]^2 ~,
\label{slice}
\end{equation}
where $ds^2_{4}$ is a K\"ahler--Einstein metric with K\"ahler form $J_4 =
\frac{1}{2} d\sigma$. It is natural to introduce the following vielbein
basis in $Y^6$, $\{dr,e^a,e^5\}$,~$a=1\dots 4$, such that, for example,
$J_4 = e^1\wedge e^2 - e^3\wedge e^4$, and the holomorphic 2-form
reads $\Omega_{4} = (e^1 + i e^2) \wedge (e^3 + i e^4)$. A set
of local complex coordinates, $\{z_1,z_2,z_3\}$, can be identified,
such that the holomorphic 3-form reads  
\begin{equation}
\Omega = e^{i\psi}\,r^2\,\Omega_{4}\,\wedge \big[\,dr+\,i\,r\,
e^5\,\big] = \frac{dz_1 \wedge dz_2 \wedge dz_3}{z_1 z_2} ~.
\label{threeform}
\end{equation}
The Killing spinors in these Sasaki--Einstein manifolds read ($\Gamma_{*}
\equiv i\,\Gamma_{x^0 x^1 x^2 x^3}$)
\begin{equation}
\epsilon\,=\,e^{-\frac{i}{2}\tilde\psi}\, r^{-\frac{\Gamma_{*}}{2}}\,\,
\Big(\,1\,+\,\frac{1}{2}\,x^{\alpha}\,\Gamma_{rx^{\alpha}}\,\,
(1\,+\,\Gamma_{*}\,)\,\Big)\,\,\eta ~,
\end{equation}
where $\Gamma_{12}\,\eta = - i \eta$, ~$\Gamma_{34}\,\eta =  i \eta$,
and $\tilde\psi$ is the angle conjugated to the $U(1)$ $R$--symmetry.

Consider a Dp--brane probe in $AdS_5 \times X^5$. The embedding can be
characterized by the set of functions $X^M(\xi^\mu)$, where $\xi^\mu$ are
the worldvolume coordinates. The supersymmetric embeddings are obtained
by imposing the condition $\Gamma_\kappa\,\epsilon = \epsilon$, where
$\epsilon$ is a Killing spinor of the background \cite{Kappa1}, and
$\Gamma_\kappa$ is a matrix that depends on the embedding \cite{Kappa2}.
Thus, $\Gamma_\kappa\,\epsilon =
\epsilon$ is a new projection giving rise to BPS equations that
determine the supersymmetric embeddings of the brane probes. It is a local
condition that must be satisfied at any point of the probe worldvolume.

\section{Dibaryon operators}

Dibaryon operators can be built for the different bifundamental fields in
the quiver gauge theory. They are pointlike objects that correspond to
supersymmetric configurations of D3--branes wrapping a three-cycle,
$\mathcal{C}_3 \subset X^5$. The homology of these manifolds allows for
several inequivalent cycles. It is important to distinguish between {\it
doublet} and {\it singlet} dibaryon operators according to the
transformation properties of the corresponding constituent chiral field
under the global $SU(2)$ symmetry.
The conformal dimension $\Delta$ of the operator dual to a D3--brane probe
wrapping $\mathcal{C}_3$, is proportional to the volume,
\begin{equation}
\Delta(\mathcal{C}_3)\,=\,\frac{\pi}{2}\, N_c\,
\frac{{\rm Vol}(\mathcal{C}_3)}{{\rm Vol}(X^5)} ~.
\label{scalingdim}
\end{equation}
We can then compute the $R$-charge of the operator since it is related to
its dimension, $R = \frac{2}{3}\Delta$. Its baryon number (in units of $N_c$)
can be obtained as the integral of the pullback of a $(2,1)$-form
\cite{HeEjKl},
\begin{equation}
{\cal B}(\mathcal{C}_3) = \pm\, i\, k_{X^5}\, \int_{\mathcal{C}_3}\,
P\Big[\,\Big(\,\frac{dr}{r} + i\, e^5\,\Big)\wedge \omega
\Big]_{\mathcal{C}_3} ~,
\label{baryonnum}
\end{equation}
where $\omega$ is a selfdual (1,1)-form satisfying $d\omega = \omega \wedge
J_4 = 0$, and $k_{X^5}$ is a constant that depends on $X^5$. Armed with these
expressions, we can extract all the relevant gauge theory information.

An exhaustive study of different D3--branes embeddings corresponding to
all possible dibaryons has been carried out for $T^{1,1}$ \cite{ArCrRa},
$Y^{p,q}$ \cite{CaEdPaZaRaVa} and $L^{a,b,c}$ \cite{CaEdRa} superconformal
field theories. This was done by demanding $\kappa$--symmetry. Besides implying
a new projection on the Killing spinor, $\Gamma_\kappa\,\epsilon = \epsilon$
also entails a set of first order BPS equations whose simplest solutions
yield a panoply of embeddings of $\mathcal{C}_3$. Compatibility with the
$AdS$ structure of the spinor implies that the D3--brane must be placed at
the center of $AdS_5$. These $1/8$ supersymmetric configurations correspond
to dibaryons in the gauge theory side. This assertion can be checked by
computing their associated $R$-charges and baryon numbers.

It is not difficult to show that more general embeddings result from the BPS
equations. Indeed, it is possible to show that these are equivalent to
Cauchy--Riemann equations for the local complex coordinates $z_1$ and $z_2$.
Then, the most general solution is given by a holomorphic embedding, $z_2 =
\mathcal{F}(z_1)$. An immediate check consists in realizing that these
generalized embeddings are calibrated,
\begin{equation}
P\Big[\,\frac{1}{2}\,J\wedge J\,\Big]_{\mathcal{D}_4}\,=\,
{\rm Vol}(\mathcal{D}_4) ~,
\end{equation}
where ${\rm Vol}(\mathcal{D}_4)\,=\,r^3\,dr\wedge {\rm Vol}(\mathcal{C}_3)$
is the volume form of the divisor.
Some of these embeddings can be understood as excitations of the dibaryons
in the case of $T^{1,1}$ \cite{ArCrRa}. However, it is important to stress
that this local condition does not always make sense globally. We have seen
examples of this feature in $Y^{p,q}$ \cite{CaEdPaZaRaVa} and $L^{a,b,c}$
\cite{CaEdRa}. \footnote{We skip all the details due to space limitations
of these Proceedings. We encourage the interested reader who is in quest
of subtleties and technicalities to look at the references
\cite{ArCrRa,CaEdPaZaRaVa,CaEdRa}.}

Excitations of a singlet dibaryon can be represented as graviton
fluctuations in the presence of the dibaryon. Instead, certain BPS excitations
of the wrapped D3--branes corresponding to doublet dibaryons can be
interpreted as a single particle state in $AdS_5$ \cite{BeHeKl}. These
excitations, roughly speaking, correspond to the insertion of a mesonic
operator $\mathcal{O}$. Thus, we have to count all possible inequivalent
(in the chiral ring) mesonic operators. \footnote{Notice that dibaryon
operators that do not reduce to those discussed here also exist in this
kind of theories \cite{Davide}. I want to thank Davide Forcella for his
instructive comments about this point.} They correspond to (short
and long) loops in the quiver diagram \cite{Benve}. The simplest ones are
operators with R-charge 2, given by short loops in the quiver. These are
the terms appearing in the tree level superpotential. They are all equivalent
in the chiral ring. Let us call its representative $\mathcal{O}_1$. It is
a spin 1 chiral operator with scaling dimension $\Delta = 3$. Its $U(1)_F$
charge vanishes. 

As for the long loops in the quiver, let us focus in the only non-trivial
case, $X^5 = Y^{p,q}$. There are two operators $\mathcal{O}_2$ and
$\mathcal{O}_3$ with spin, respectively, $\frac{p+q}{2}$ and $\frac{p-q}{2}$.
They have a nonvanishing $U(1)_F$ charge. These are the building blocks of
other chiral operators, ${\cal O}=\prod_{i=1}^3{\cal O}_{i}^{\,\,n_i}$.
The spectrum of fluctuations of these dibaryons along the transverse $S^2$
can be worked out \cite{CaEdPaZaRaVa}. The action for the D3--brane should
be expanded around the static configuration, $g = g_{(0)} + \delta g$. At
quadratic order, it is possible to identify ground state solutions with BPS
operators. Their conformal dimensions can be read off, and the spectrum
can be shown to coincide with the mesonic chiral operator quantum numbers.

\section{The baryon vertex}

For a D5--brane that wraps the whole $X^5$ space, the flux of the RR
$F^{(5)}$--form acts as a source for the electric worldvolume gauge field
which, in turn, gives rise to a bundle of fundamental strings emanating
from the D--brane. The probe action must include the worldvolume field $F$
in both the Dirac--Born--Infeld (DBI) and Wess--Zumino (WZ) terms:
\begin{equation}
S\,=\,-T_5\,\int d^6\xi\,\sqrt{-\det (g+F)}\,+\,
T_5\int d^6\xi \,\,\,A\wedge F^{(5)} ~.
\label{Sbaryonv}
\end{equation}
We were unable to find a supersymmetric configuration. From the point of
view of $\kappa$--symmetry, it turns out that the new projection,
$\Gamma_\kappa\,\epsilon = \Gamma_{x^0r}\,\epsilon^* = \epsilon$, which,
as expected, corresponds to fundamental strings in the radial direction,
cannot be imposed on the Killing spinors. Besides, it is also possible to
show that, from the point of view of the worldvolume theory, there are no
solitons saturating a Bogomol'nyi bound. Thus, we conclude from this
incompatibility argument that the baryon vertex configuration breaks
completely the supersymmetry of the $AdS_5 \times X^5$ background. 

\section{Fractional brane}

Consider a D5--brane probe that wraps a two-dimensional submanifold $L_2$
of $X^5$ and is a codimension one object in $AdS_5$. In the field theory side,
this is the kind of brane that represents a domain wall across which the
rank of the gauge groups jumps. The upshot of the detailed analysis
accomplished in \cite{ArCrRa,CaEdPaZaRaVa,CaEdRa} is as follows. We have
shown that the cone $\mathcal{L}_3 = \mathcal{C}(L_2)$, is calibrated. Indeed,
the holomorphic $(3,0)$ form $\Omega$ of $\mathcal{C}(X^5)$ --see
eq.(\ref{threeform})--, can be naturally used to calibrate such submanifolds:
$\mathcal{L}_3$ is called a special Lagrangian submanifold of $\mathcal{C}(X^5)$
if the pullback of $\Omega$ to $\mathcal{L}_3$ is, up to a constant phase
$\lambda$, equal to its volume,
\begin{equation}
P\big[\,\Omega\,\big]_{\mathcal{L}_3}\,=\,e^{i\lambda}\,
{\rm Vol}\,(\mathcal{L}_3) ~.
\label{D5calibration}
\end{equation}
The fractional brane can be
also understood as a BPS worldvolume soliton. This arises from the Hamiltonian
density resulting from the DBI action, since it can be written as a sum of
squares in such a way that it becomes minimum when a set of BPS differential
equations are satisfied. Not surprisingly, they agree with those obtained
from the $\kappa$--symmetry approach.

\section{Flavor D7--branes}

The D7-branes which fill the four Minkowski spacetime dimensions and extend
along some holographic non-compact direction can be potentially used as flavor
branes, {\it i.e.} as branes whose fluctuations can be identified with the
dynamical mesons of the gauge theory \cite{KaKa}. The ansatz we adopt for
the worldvolume coordinates is $\xi^\mu=(x^\alpha,\theta^\beta)$, and we consider
embeddings of the form $r = r(\theta^\beta)$ and $\psi = \psi(\theta^\beta)$.
In order to implement $\Gamma_{\kappa}\,\epsilon\,=\,\epsilon$, we
shall require that the spinor $\epsilon$ is an eigenvector of the matrix
$\Gamma_*$ defined above.
These configurations preserve the four ordinary supersymmetries of the
background. By means of the $\kappa$--symmetry technique, it is possible
to show that a generic configuration can be nicely written as a holomorphic
embedding \cite{ArCrRa,CaEdPaZaRaVa,CaEdRa}
\begin{equation}
z_1^{m_1}\,z_2^{m_2}\,z_3^{m_3}\,=\,{\rm constant}\,\,,
\label{D7polynomial}
\end{equation}
where the $m_i$'s are constants and $m_3\not= 0$. These configurations
seem to be the relevant ones to introduce matter in the fundamental
representation \cite{Flavors}.

The identification of supersymmetric 4-cycles that a D7--brane can wrap
also matters in cosmological models where inflation is produced by the motion
of a D3--brane in a warped throat. The potential ruling this motion is
actually sensitive to the specific embedding of the wrapped D7--brane 
\cite{DyGuGuMa}.

\section{Further configurations}

Another cases of interest include a non-supersymmetric (still stable) probe
D3--brane extended along one gauge theory direction and wrapping a 2-cycle
(a {\it fat string} from the gauge theory point of view), a D5--brane that
extends infinitely in the holographic direction (a {\it defect CFT} that
preserves four supersymmetries), and a D7--brane wrapping the whole $X^5$
space, and being codimension two in $AdS_5$ (a supersymmetric {\it string}).
In the D5--brane configuration, we have also turned on a worldvolume flux
and found that it leads to a bending of the profile of the wall.
Details can be found in the original references
\cite{ArCrRa,CaEdPaZaRaVa,CaEdRa}.

\begin{acknowledgement}
I am pleased to thank Felipe Canoura, Leo Pando Zayas, Alfonso Ramallo
and Diana Vaman for delightful collaborations leading to the
results presented in this talk. This work was supported in part by MCyT
and FEDER (grant FPA2005-00188), Xunta de Galicia (Conseller\'\i a de
Educaci\'on and grant PGIDIT06PXIB206185PR), and the EC Commission
(grant MRTN-CT-2004-005104). The author is a {\it Ram\'on y Cajal}
Research Fellow.
Institutional support to the Centro de Estudios Cient\'\i ficos (CECS) from
Empresas CMPC is gratefully acknowledged. CECS is funded in part by grants
from Millennium Science Initiative, Fundaci\'on Andes and the
Tinker Foundation.
\end{acknowledgement}


\end{document}